\documentclass[pra,aps,longbibliography,showpacs,groupedaddress,superscriptaddress,floatfix,twocolumn,toc=flat]{revtex4-1}

\usepackage{amsfonts,amsmath,amssymb,stmaryrd}

\usepackage[utf8]{inputenc}
\usepackage[table]{xcolor}
\usepackage{bbold, bm}
\usepackage{graphicx}
\usepackage{array,multirow}
\usepackage{algorithm}
\usepackage{algpseudocode}
\algrenewcommand\algorithmicrequire{\textbf{Input:}}
\algrenewcommand\algorithmicensure{\textbf{Output:}}
\usepackage{boldline}
\usepackage{tabularx}
\usepackage{tikz}
\usepackage{placeins}
\usepackage{mathrsfs}
\pdfmapline{-dummy LMRoman10-Regular} 
\pdfmapline{-dummy LMRoman10-Bold} 
\pdfmapline{-dummy LMRoman10-Italic} 
\pdfmapline{-dummy LMRoman10-BoldItalic} 
\pdfmapline{-dummy LMRoman12-Regular} 
\pdfmapline{-dummy LMRoman12-Bold} 
\pdfmapline{-dummy LMRoman12-Italic} 
\pdfmapline{-dummy LMRoman12-BoldItalic} 

\usepackage{ctable}
\usepackage{epsfig}
\usepackage[english, status=final]{fixme}
\fxsetup{marginface=\tiny}
\fxusetheme{color}    

\usepackage[colorlinks=true,linkcolor=blue,urlcolor=blue,citecolor=blue]{hyperref}

\newcommand{\bra}[1]{\langle#1|}
\newcommand{\ket}[1]{|#1\rangle}

\renewcommand{\k}{k}
\renewcommand{\i}{i}
\renewcommand{\j}{j}

\makeatletter
\def\maketitle{
\@author@finish
\title@column\titleblock@produce
\suppressfloats[t]}
\makeatother

\AtBeginDocument{%
    \newwrite\bibnotes
    \def\bibnotesext{Notes.bib}
    \immediate\openout\bibnotes=\jobname\bibnotesext
    \immediate\write\bibnotes{@CONTROL{REVTEX41Control}}
    \immediate\write\bibnotes{@CONTROL{%
    apsrev41Control,author="08",editor="1",pages="1",title="0",year="1"}}
     \if@filesw
     \immediate\write\@auxout{\string\citation{apsrev41Control}}%
    \fi
}%

\begin{document}

\title{Emergent spinon-holon Feshbach resonance in a doped Majumdar-Ghosh model}

\author{Simon M. Linsel}
\email{simon.linsel@lmu.de}
\affiliation{Department of Physics and Arnold Sommerfeld Center for Theoretical Physics (ASC), Ludwig-Maximilians-Universit\"at M\"unchen, Theresienstr. 37, M\"unchen D-80333, Germany}
\affiliation{Munich Center for Quantum Science and Technology (MCQST), Schellingstr. 4, D-80799 M\"unchen, Germany}
\affiliation{Department of Physics, Harvard University, Cambridge MA 02138, USA}


\author{Ulrich Schollwöck}
\affiliation{Department of Physics and Arnold Sommerfeld Center for Theoretical Physics (ASC), Ludwig-Maximilians-Universit\"at M\"unchen, Theresienstr. 37, M\"unchen D-80333, Germany}
\affiliation{Munich Center for Quantum Science and Technology (MCQST), Schellingstr. 4, D-80799 M\"unchen, Germany}

\author{Annabelle Bohrdt}
\affiliation{Munich Center for Quantum Science and Technology (MCQST), Schellingstr. 4, D-80799 M\"unchen, Germany}
\affiliation{Institute of Theoretical Physics, University of Regensburg, D-93053, Germany}

\author{Fabian Grusdt}
\email{fabian.grusdt@lmu.de}
\affiliation{Department of Physics and Arnold Sommerfeld Center for Theoretical Physics (ASC), Ludwig-Maximilians-Universit\"at M\"unchen, Theresienstr. 37, M\"unchen D-80333, Germany}
\affiliation{Munich Center for Quantum Science and Technology (MCQST), Schellingstr. 4, D-80799 M\"unchen, Germany}

\date{\today}
\begin{abstract}
Experimental and numerical spectroscopy have revealed rich physics in antiferromagnets, in particular in frustrated and doped systems. The Majumdar-Ghosh (MG) model has an analytically known spin-disordered ground state of dimerized singlets as a result of magnetic frustration. Here we study the single-hole angle-resolved photoemission spectrum (ARPES) of a doped MG model, where we introduce a spin-hole interaction that is experimentally accessible with ultracold molecules. We report a bound spinon-holon ground state and clear signatures of a spinon-holon molecule state and polarons in the ARPES spectrum at different magnetizations. Moreover, we find signatures of an emergent Feshbach resonance with tunable interactions associated with the unbinding of the spinon and the holon. Our results provide new insights into the physics of dopants in frustrated $t$-$J$ models and establish the latter as a new platform for studies of emergent few-body phenomena.
\end{abstract}
\maketitle


\section{Introduction}

The resonating valence bond theory (RVB), developed by Anderson and Fazekas \cite{Anderson1973, Fazekas1974}, describes a quantum spin liquid (QSL) on a triangular lattice with featureless constituents: holons and spinons. Historically, RVB was proposed to describe high-temperature superconductivity in the 2D Hubbard model \cite{Anderson1987, Baskaran1987, Baskaran1988}. While the RVB paradigm is still widely applied for describing spin liquids, in the context of doped antiferromagnets theories with confined phases or non-trivial constituents have emerged in recent years. A prominent example are fractionalized Fermi liquids (FL*) \cite{Senthil2003, Senthil2004}, often studied in the context of doped quantum dimer models \cite{Punk2015}. This parton picture is in line with microscopic studies of doped holes \cite{Bohrdt2021} and hole pairs \cite{Bohrdt2023} in $t$-$J$ and Hubbard models.

Feshbach resonances have originally been introduced in the context of particle physics, where slow-moving colliding particles undergo resonant scattering \cite{Feshbach1962}. Since then, Feshbach resonances have been widely used to realize tunable interactions in cold-atom experiments \cite{Courteille1998, Inouye1998, Bloch2008, Chin2010} and 2D semiconductors \cite{Sidler2017, Schwartz2021}. Recently, Feshbach resonances have further been proposed as a possible pairing mechanism for high-temperature superconductivity in cuprates \cite{Squire2010, Homeier2023, Homeier2024}.

In this Article, we report Feshbach-like resonant interactions upon tuning across the spinon/holon unbinding in a paradigmatic doped frustrated quantum magnet. We study the doped Majumdar-Ghosh model \cite{Majumdar1969a, Majumdar1969b} extended by spin-hole interactions that can be realized e.g. by ultracold polar molecules \cite{Gorshkov2011, Carroll2024}, serving as a toy model relevant for other settings featuring spinon-holon bound states. The resonant spinon-holon interactions we reveal are directly probed by varying the density of unpaired spinons. Using matrix product states (MPS), we study the ground state properties and calculate the single-hole ARPES spectrum. In addition to the Feshbach-like resonance, we find a rich set of emergent few-body states realized in the doped MG model. Our results have possible implications for the physics of cuprates in the pseudogap regime.



\section{Doped Majumdar-Ghosh model}

\begin{figure}[t]
\includegraphics[width=0.49\textwidth]{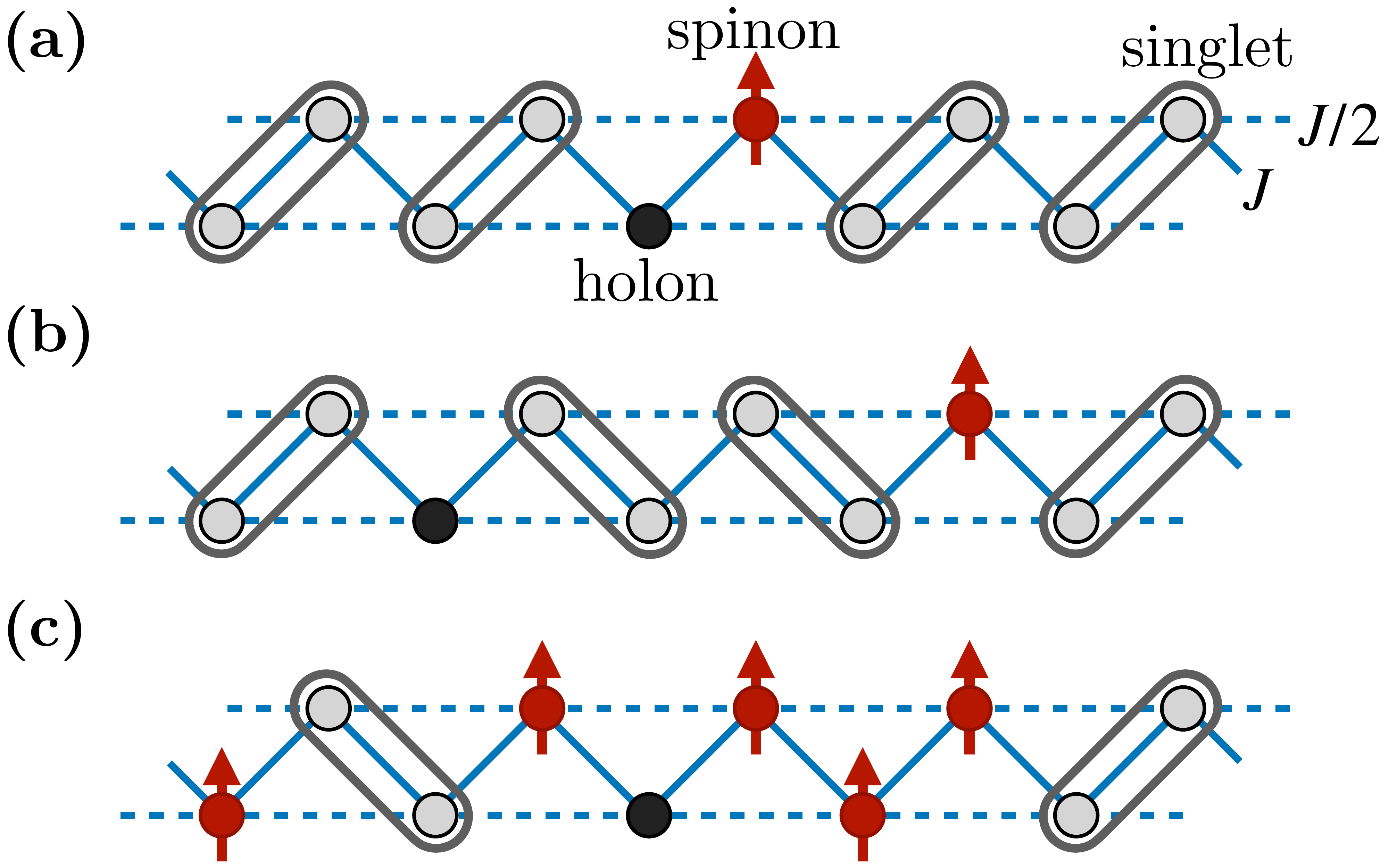}
\caption{\textbf{Spinon-holon bound state in the doped Majumdar-Ghosh model.} The bare Majumdar-Ghosh model features a ground state of dimerized spin singlets. We illustrate \textbf{(a)} a bound and \textbf{(b)} an unbound spinon-holon state in the Majumdar-Ghosh model with one hole. \textbf{(c)} We show a Majumdar-Ghosh model with one hole and high magnetization. The unpaired spinons form a Luttinger liquid.}
\label{fig1}
\end{figure}

We study the frustrated Hamiltonian
\begin{align}
    \hat{\mathcal{H}} = &-t \sum_{ \langle \i,\j \rangle, \sigma } \hat{\mathcal{P}}_{\mathrm{GW}} \bigl[ \hat{c}_{\i,\sigma}^{\dagger} \hat{c}_{\j, \sigma} + \mathrm{H.c.} \bigr] \hat{\mathcal{P}}_{\mathrm{GW}} + J \sum_{\j=1}^{N} \hat{\boldsymbol{S}}_{\j} \cdot \hat{\boldsymbol{S}}_{\j+1} \nonumber \\ 
    &+ \frac{J}{2} \sum_{\j=1}^{N} \hat{\boldsymbol{S}}_{\j} \cdot \hat{\boldsymbol{S}}_{\j+2} - g \sum_{\langle \i,\j \rangle} \bigl[ \hat{n}^{\mathrm{h}}_{\i} \hat{S}^z_{\j} + \mathrm{H.c.} \bigr] \label{eq:eMG}
\end{align}
on the triangular ladder, see Fig.~\ref{fig1}, where $\hat{c}_{\j, \sigma}$ is a fermionic annihilation operator on site $\j$ with spin $\sigma \in \{ \uparrow, \downarrow \}$, $\hat{\boldsymbol{S}}_{\j}$ is the spin operator and $\hat{n}^{\mathrm{h}}_{\j} = \prod_{\sigma} (1 - \hat{c}_{\j, \sigma}^\dagger \hat{c}_{\j, \sigma})$ is the hole density. It includes the bare Majumdar-Ghosh model featuring Heisenberg interactions between nearest neighbors (NN) and next-nearest neighbors (NNN) $\propto J$. We allow NN hopping of fermions $\propto t$; here $\hat{\mathcal{P}}_{\mathrm{GW}}$ denotes the Gutzwiller projector on states with no more than one fermion per site. In addition, we add a spin-hole interaction $\propto g$ that can be experimentally realized using e.g. ultracold polar molecules \cite{Gorshkov2011, Carroll2024}, ultracold atoms \cite{Jepsen2021}, or in Rydberg tweezer arrays \cite{Homeier2024_2}. We set $t=J$; this choice is not particularly special.

\begin{figure}[t]
\includegraphics[width=0.49\textwidth]{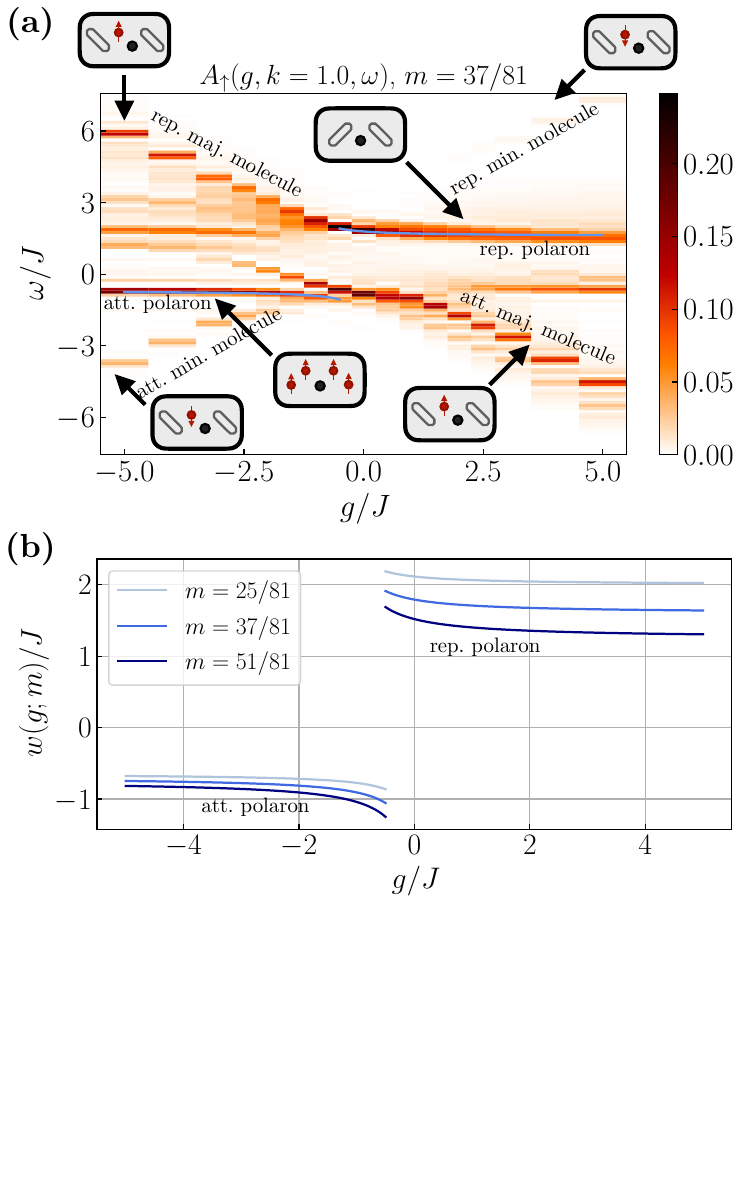}
\caption{\textbf{Emergent Feshbach-like interaction.} We perform microscopic MPS simulations of the single-hole ARPES spectrum Eq.~(\ref{eq:arpes}) of Hamiltonian~(\ref{eq:eMG}). Details of the MPS calculations can be found in the Appendix Sec.~\ref{app:convergence}. \textbf{(a)} We show the one-hole majority ARPES spectrum $A_{\uparrow}$ (i.e. we remove an $\uparrow$-spinon and probe the time evolution of the resulting hole) at fixed momentum $ka=1.0$ (corresponding approximately to the GS of the one-hole dispersion; $a$ is the lattice constant) for $L=81$ and magnetization $m=37/81$. We find signatures of attractive/repulsive Fermi polarons. We probe the Feshbach resonance by tuning the density of unpaired spinons ($m$) and extracting the peak positions $\omega(g;m)$ of the polaron branches, see \textbf{(b)}. The polaron branches repel each other for increasing $m$, signaling a resonantly enhanced interaction.}
\label{fig2}
\end{figure}

We microscopically simulate the system using MPS and apply the density-matrix renormalization group (DMRG) \cite{Schollwoeck2011} to calculate the ground state (GS). Time evolutions of the MPS are obtained using generalized subspace expansion (GSE) \cite{Yang2020} for the first few time steps and then using the two-site time-dependent variational principle (TDVP2) \cite{Paeckel2019}. In addition, we use linear extrapolation to improve the quality of the ARPES spectrum \cite{Barthel2009}. We enforce a global $U(1) \otimes U(1)$ symmetry of the hole number and the magnetization $m = 2 \langle \sum_{\j} \hat{S}^z_{\j} \rangle / L$. The magnetization is always positive in this Article. We use the \textsc{SyTen} toolkit \cite{Syten}.

\subsection{Spinon-holon Feshbach resonance}

The undoped Majumdar-Ghosh model, i.e. Hamiltonian~(\ref{eq:eMG}) at $\hat{n}_{\j}^{\mathrm{h}} = 0$, features a GS of dimerized singlets, see Fig.~\ref{fig1}. We dope one hole into the system and set the magnetization to a small but non-zero value, i.e. we have some unpaired spinons in the system that are not bound in a singlet. Consequently, the GS of this doped and magnetized system is translationally invariant. By tuning the microscopic spin-hole interaction $\propto g$, we can realize both an unbound and a bound regime, in which a spinon-holon bound state forms. The existence of the bound state itself is very natural and we will study the microscopic details later; for now, it is only important that its existence can be controlled via $g$.

This brings us to the central idea put forward in this Article: We propose that the \textit{unbinding transition} of the spinon-holon bound state is associated with a \textit{Feshbach-type resonance} with resonantly enhanced scattering. 

To reveal signatures of such resonantly enhanced spinon-holon interactions, we consider a system at finite magnetization. For the moment, let us \textit{assume} a Feshbach-like resonance, in which the spinon and the holon undergo two-body scattering: the interaction energy near the transition is given by
\begin{align} \label{eq:mean_field}
    \hat{\mathcal{H}}_{\mathrm{int}} \sim -g_\mathrm{eff}(g) \hat{n}^{\mathrm{h}} \langle \hat{n}^S \rangle,
\end{align}
where $g_\mathrm{eff}(g)$ is the resonantly enhanced effective interaction that depends on the bare coupling $g$ and $\hat{n}^S$ is the unpaired spinon density. Here we employed a mean-field ansatz, replacing the unpaired spinon density by its average, i.e. $\hat{n}^S \to \langle \hat{n}^S \rangle = m$. Thus increasing $m$ leads to a stronger interaction energy shift in the system. This situation resembles a 1D Fermi polaron problem \cite{Massignan2014}, where the holon corresponds to an impurity. The unpaired spinons around the holon form a Luttinger liquid and ``dress'' the holon in the following sense: The attractive (repulsive) polaron is a holon that is surrounded by $\uparrow$-spinons (singlets). Since the unpaired spinons are mutually hard-core, we expect the polarons to resemble 1D Fermi polarons.

\begin{figure}[t]
\includegraphics[width=0.49\textwidth]{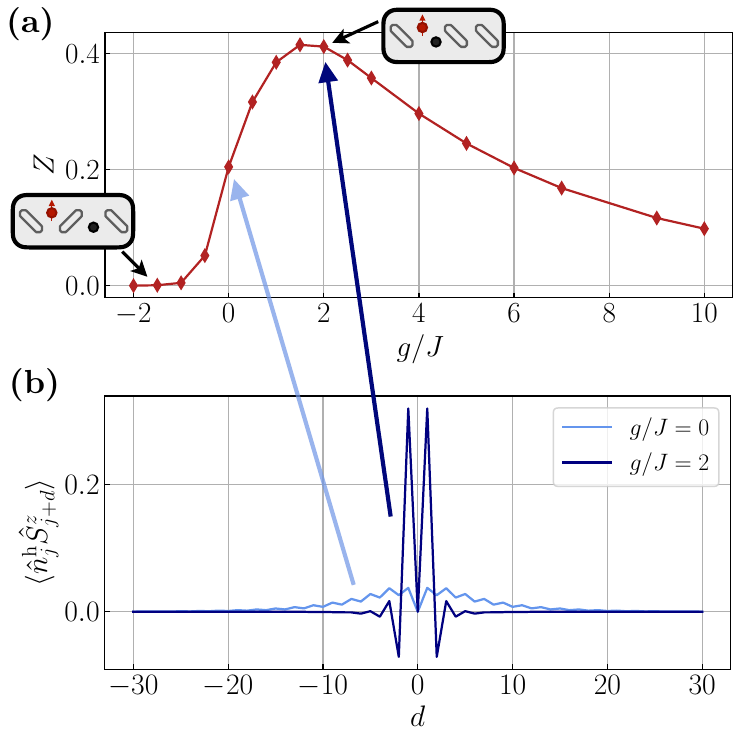}
\caption{\textbf{Spinon-holon bound state signatures.} We show the effect of tuning the interaction $g$ in the DMRG ground state of Hamiltonian~(\ref{eq:eMG}). \textbf{(a)} We show the quasiparticle weight $Z(g)$. We observe a maximum around $g/J = 2$, where sizable values $Z \sim 0.4$ are realized. \textbf{(b)} We show the spin-hole correlator $\langle \hat{n}_{\j}^{\mathrm{h}} \hat{S}^z_{\j + d} \rangle$ averaged over the system. For $g=0(2)$, we observe a bound spinon-holon state which extends over approximately 20 (a few) sites, i.e. the size of the bound state decreases with increasing $Z$.}
\label{fig3}
\end{figure}

In order to extract $g_\mathrm{eff}(g)$, we start by tuning $g, m$ and searching for Fermi polaron signatures in the one-hole ARPES spectrum
\begin{align} \label{eq:arpes}
    A_{\sigma}(\k, \omega) = \frac{1}{2 \pi} \mathrm{Re} \int_{- \infty}^{\infty} \mathrm{d}t \, e^{i \omega t} \bra{\psi_0} e^{i \hat{\mathcal{H}} t} \hat{c}_{\k, \sigma}^{\dagger} e^{-i \hat{\mathcal{H}} t} \hat{c}_{\k, \sigma} \ket{\psi_0},
\end{align}
where $\ket{\psi_0}$ is the GS without holes. We calculate the full one-hole ARPES spectrum, thus going beyond mean-field theory. In addition to Fermi polaron-like branches, whose energy we will model by Eq.~(\ref{eq:mean_field}) to extract $g_\mathrm{eff}(g)$, the spectrum reveals a rich set of few-body states, e.g. molecular spinon-holon states. First, we will focus on the polaron branches as they are immediately relevant to establish proof of the Feshbach resonance; however, we will discuss the other branches later. 

We fix $ka=1.0$ (corresponding approximately to the GS of the one-hole dispersion; $a$ is the lattice constant) and plot the majority ($\sigma = \; \uparrow$) ARPES spectrum with respect to $g$ for different magnetizations $m$; our result for $m=37/81$ is shown in Fig.~\ref{fig2}a. Indeed, we identify two branches with attractive and repulsive Fermi polaron characters, respectively. The attractive Fermi polaron has a negative energy compared to the zero-hole GS energy $E_0 \equiv 0$. It has a particularly large spectral weight in the \textit{repulsive} region ($g<0$) of the bare coupling $g$. This is in line with the Feshbach picture: A repulsive bare interaction corresponds to a strong resonantly enhanced effective attractive interaction and vice versa, giving rise to the existence of the two polaron branches. 

As a side note, the attractive polaron might still exist deep in the attractive region $g>0$ while the repulsive polaron will typically decay because of its high energy compared to the GS. The repulsive Fermi polaron has a positive energy and a large spectral weight for attractive bare interactions ($g>0$). In Fig.~\ref{fig2}a we only fit the peak positions (blue lines) in the parts of the two polaron branches with high spectral weight.

We repeat these calculations for different magnetizations and show the fitted peak positions of the polaron branches in Fig.~\ref{fig2}b. We clearly observe a \textit{repulsion} of the attractive/repulsive branches with increasing $m$, as expected from Eq.~(\ref{eq:mean_field}), thus strongly suggesting a resonantly enhanced Feshbach interaction at $g_{\mathrm{c}} \approx -0.5$ ($J \equiv 1$). We find that the fitted peak positions of the polaron branches indeed satisfy Eq.~(\ref{eq:mean_field}) with $g_\mathrm{eff}(g) \sim g_\mathrm{bg} - 1/g$ and $g_\mathrm{eff}(g_{\mathrm{c}}) = 2.7 \times |g_{\mathrm{c}}|$, i.e. a \textit{weak}, \textit{repulsive} microscopic $g$ corresponds to a \textit{strong}, \textit{attractive} $g_\mathrm{eff}(g)$, see Appendix Sec.~\ref{app:arpes_spectra}. 

\begin{figure}[t]
\includegraphics[width=0.49\textwidth]{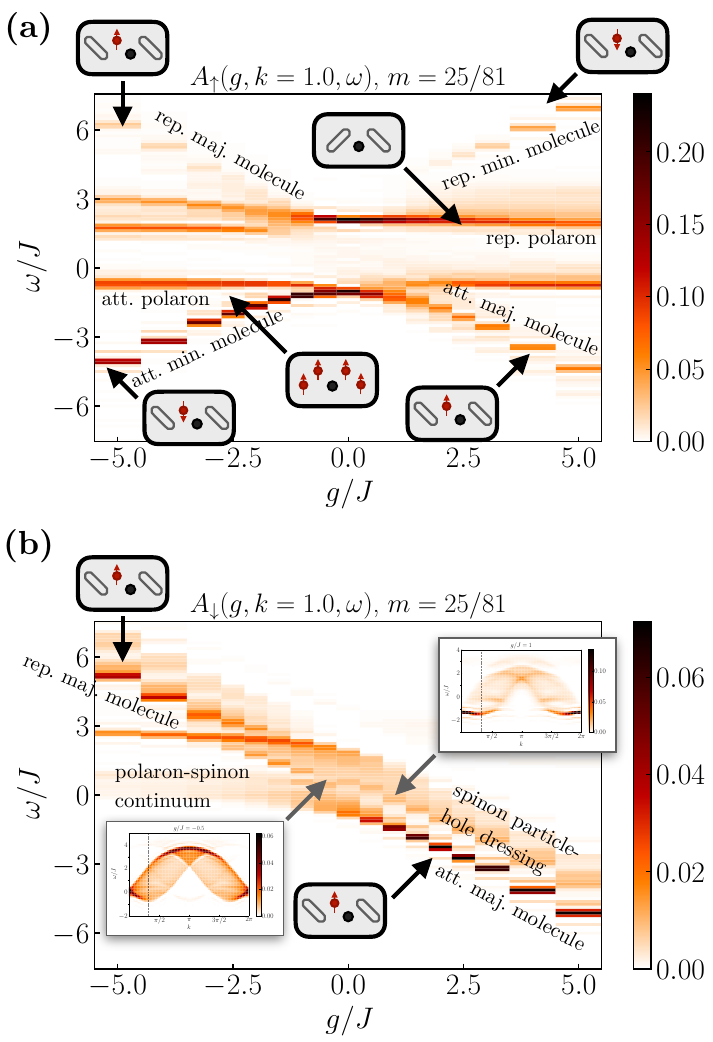}
\caption{\textbf{Majority and minority ARPES spectrum.} We perform microscopic MPS simulations of the single-hole ARPES spectrum Eq.~(\ref{eq:arpes}) of Hamiltonian~(\ref{eq:eMG}) at fixed momentum $k=1.0$ for $L=81$ and magnetization $m=25/81$. We tune the spin-hole interaction $\propto g$ and show the ARPES spectrum at different interactions. ARPES spectra for other $m$ can be found in the Appendix Sec.~\ref{app:arpes_spectra}. \textbf{(a)} We show the majority spectrum $A_{\uparrow}$. The linear branches represent the majority/minority spinon-holon molecule states, which can be attractively/repulsively bound, respectively. \textbf{(b)} We show the minority spectrum $A_{\downarrow}$. The minority molecule is not visible due to vanishing spectral weight. In addition to the molecule branches, we observe a polaron-spinon continuum signaled by a cosine dispersion of the momentum-dependent ARPES (see lower left inset), and spinon-particle hole dressing signaled by a continuum of states above the attractive majority molecule ground state (see upper right inset).}
\label{fig4}
\end{figure}

We will discuss the microscopic details of the spinon-holon bound state next, after making one more remark: the Feshbach resonance is connected to the unbinding of a spinon-holon pair, our ARPES calculation is just a way to \textit{probe} it. In fact, we argue that Feshbach resonances might be more generally prevalent in scenarios where partons form bound states, e.g. in the context of FL-FL* transitions \cite{Senthil2003, Senthil2004}: in the FL* phase deconfined spinons and holons exist but the relevant charge carriers at low energies are constituted by bound pairs of spinons and holons, yielding electron-like quantum numbers.

\subsection{Free molecule limit}

We study the microscopic structure of the bound spinon-holon state at minimal magnetization, i.e. with only one hole and one unpaired spinon in the system. To study the effect of tuning the spin-hole interaction $\propto g$, we calculate the QP weight
\begin{align} \label{eq:qp_weight}
    Z = \sum_{\j} | \bra{\psi_0^{\mathrm{1h}}} \hat{c}_{\j, \downarrow} \ket{\psi_0^{\mathrm{0h}}} |^2,
\end{align}
where $\ket{\psi_0^{\mathrm{0(1)h}}}$ is the DMRG GS with zero (one) holes at $m=0(1/2)$. It includes the MPS overlap between the one-hole DMRG GS and the undoped GS in which we put a spinon directly next to a holon (by removing one spin from a singlet). Thus $Z$ probes the existence of a spinon-holon bound state. Note that in contrast to the ARPES simulations, here we use an even system size of $L=80$. We perform a scan over $-2 \leq g/J \leq 10$ and plot $Z$ in Fig.~\ref{fig3}a. $Z$ is zero for $g/J \ll -1$ and features a maximum around $g/J = 2$, where sizable values $Z \sim 0.4$ are realized. This confirms the formation of a spinon-holon bound state for $g>0$. Interestingly, the bare interaction $g$ at the approximate location of the resonance coincides with the transition from the unbound to the bound regime in the free molecule limit, see Fig.~\ref{fig3}a which is consistent with the bare interaction $g$ driving the transition. 

To study the microscopic structure of the bound state further, we calculate the spin-hole correlator $\langle \hat{n}_{\j}^{\mathrm{h}} \hat{S}^z_{\j + d} \rangle$ for $g/J=0$ and $2$. For $g/J=0$, we observe a large bound state that extends over approximately 20 lattice sites, still well below the system size $L=80$. The size of the bound state decreases significantly down to a few lattice sites for $g/J=2$. Thus, the existence and the size of a bound spinon-holon state are directly tunable by $g/J$. 

\subsection{Emergent few-body physics}

The calculated ARPES spectra feature not only Fermi polarons but also a rich set of further branches, some of which we will identify here. We show the majority ($\sigma = \; \uparrow$) ARPES spectrum for different spin-hole interactions $\propto g$ at $m=25/81$ in Fig.~\ref{fig4}a. In addition to the Fermi polarons, we identify 4 branches where a spinon and the holon form a molecular state, signaled by a linear dependence of the energy $E \sim \pm |g|$. We call the state where the holon binds to a $\uparrow$-spinon ($\downarrow$-spinon) a majority (minority) molecule. Both molecule types can be either attractively (negative energy) or repulsively (positive energy) bound. A repulsively bound molecule is an excited state that can be long-lived due to a lack of low-order resonant decay processes.

We show the minority ($\sigma = \; \downarrow$) ARPES spectrum for different spin-hole interactions $\propto g$ at $m=25/81$ in Fig.~\ref{fig4}b. In addition to the attractive and repulsive majority molecule, we observe a polaron-spinon continuum enclosed by a cosine dispersion in the momentum-dependent ARPES spectrum at fixed $g/J$, see inset of Fig.~\ref{fig4}b. Further, we observe spinon particle-hole dressing as a continuum of excited states above the attractive majority molecule, where the interaction between the impurity (i.e. the holon) with the Luttinger liquid results in particle-hole excitations (i.e. collective excitations of spinons). Attentive readers may have noticed that the minority molecule and the Fermi polarons are not visible in the minority spectrum. This is due to their low spectral weight as a result of vanishing wave function overlaps in the ARPES spectrum, as we discuss further in the Appendix Sec.~\ref{app:arpes_spectra}. More detailed studies of the molecular branches and the other few-body excitations not discussed here are tasks left for future research.


\section{Conclusion}

We have investigated a doped Majumdar-Ghosh model as a paradigmatic frustrated quantum magnet. The model can be experimentally implemented using e.g. ultracold dipolar molecules \cite{Gorshkov2011, Carroll2024}, and for $g=0$ using ultracold fermions in quantum gas microscopes \cite{Bohrdt2021_1, Tarruell2018, Esslinger2010, Prichard2024, Xu2023, Yang2021}. Using ARPES based on MPS, we have found signatures of a spinon-holon Feshbach-like resonance which can be probed by tuning the density of unpaired spinons (i.e. magnetization). We have also found emergent few-body physics in the ARPES spectra, including different (partly repulsively bound) molecular branches of a spinon and a holon.

Our results suggest wider applicability of this emergent Feshbach-like resonance to study the physics of superconducting phases or parton unbinding. It is potentially relevant to clarify the physics of doped quantum spin liquids and has potential applications in the context of heavy electrons \cite{Loehneysen1994, Loehneysen1996} and cuprates.

\section*{Acknowledgments}

We thank T. Blatz, M. Grundner, L. Homeier, M. Kebri\v c, N. Mostaan, S. Paeckel, F. Palm, H. Schlömer and Y. Tan for fruitful discussions.
This research was funded by the European Research Council (ERC) under the European Union’s Horizon 2020 research and innovation program (Grant Agreement no 948141) — ERC Starting Grant SimUcQuam, and by the Deutsche Forschungsgemeinschaft (DFG, German Research Foundation) under Germany's Excellence Strategy -- EXC-2111 -- 390814868.

\section*{Appendix}
\setcounter{section}{0}

\section{Calculating the ARPES spectrum  \& MPS convergence}
\label{app:convergence}

We perform microscopic MPS simulations of Hamiltonian~(\ref{eq:eMG}) for magnetization $m=13/81,25/81,37/81,51/81$ and spin-hole interaction $-5 \leq g/J \leq 5$. Note that in our convention $m = 2 \langle \sum_{\j} \hat{S}^z_{\j} \rangle / L > 0$, i.e. each spin contributes $\pm 1/L$ to $m$. Our procedure is similar to \cite{Bohrdt2020} but here we use the \textsc{SyTen} toolkit \cite{Syten}. To probe signatures of Feshbach-like resonances, we calculate the majority ($\sigma = \; \uparrow$) and minority ($\sigma = \; \downarrow$) ARPES spectrum Eq.~(\ref{eq:arpes}):
\begin{align*} 
    A_{\sigma}(\k, \omega) = \frac{1}{2 \pi} \mathrm{Re} \int_{- \infty}^{\infty} \mathrm{d}t \, e^{i \omega t} A_{\sigma}(\k, t),
\end{align*}
where
\begin{align*} 
    A_{\sigma}(\k, t) = \frac{1}{L} \sum_{\i, \j} e^{-i \k ( \i - \j )} \bra{\psi_0} e^{i \hat{\mathcal{H}} t} \hat{c}_{\j, \sigma}^{\dagger} e^{-i \hat{\mathcal{H}} t} \hat{c}_{\i, \sigma} \ket{\psi_0}.
\end{align*}
First, we calculate the MPS groundstate (GS) $\ket{\psi_0}$ without holes using DMRG with bond dimension $\chi=1000$. The GS is well converged with variance $\langle \hat{\mathcal{H}}^2 \rangle - \langle \hat{\mathcal{H}} \rangle^2 < 10^{-11}$. 

\begin{figure}[t]
    \centering
    \includegraphics[width = 0.49 \textwidth]{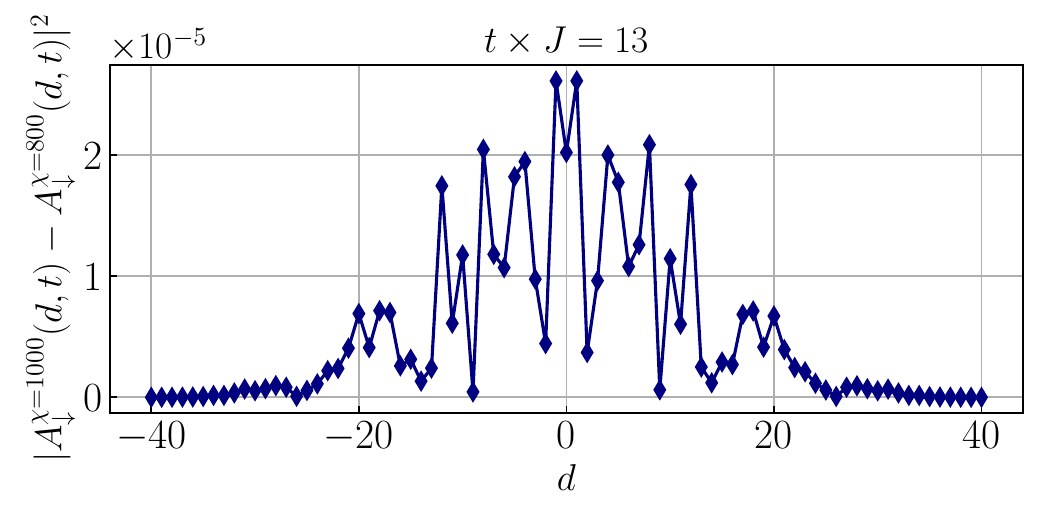}
    \caption{\textbf{MPS time evolution convergence.} We show the difference of the minority Green's function $A_{\downarrow}(d, t) = \langle e^{i \hat{\mathcal{H}} t} \, \hat{c}^{\dagger}_{L/2 + d} \, e^{-i \hat{\mathcal{H}} t} \, \hat{c}_{L/2} \rangle$ evaluated at bond dimensions $\chi=800$ and $1000$, at time $t=13/J$. We exponentially suppress times after $t=40/3J$ with a Gaussian $w(t) = \exp(-2(t/t_0)^2)$ where $t_0 = 20/3J$. The difference is $\sim 10^{-5}$, thus the time evolution is well converged.}
    \label{fig:convergence}
\end{figure}

Then, we dope a hole into the system and calculate the time evolution up to times $t = 20/J$ using generalized subspace expansion (GSE) \cite{Yang2020} for the first few time steps and then using the two-site time-dependent variational principle (TDVP2) \cite{Paeckel2019} for later times. We choose the time step $\Delta t = 0.02/J$. In Fig.~\ref{fig:convergence}, we compare the minority Green's function $\langle e^{i \hat{\mathcal{H}} t} \, \hat{c}^{\dagger}_{L/2 + d} \, e^{-i \hat{\mathcal{H}} t} \, \hat{c}_{L/2} \rangle$ for bond dimensions $\chi=800$ and $1000$, at time $t=13/J$. The difference is $\sim 10^{-5}$, thus the time evolution is well converged. We calculate the MPS overlap with the original GS $\ket{\psi_0}$. 

Next, we perform a Fourier transform into momentum space to obtain $A_{\sigma}(\k, t)$. To prevent unphysical oscillations in the time Fourier transform resulting from a cutoff at $t=20/J$, we apply linear extrapolation \cite{Barthel2009}. As a tradeoff between the error of the TDVP2 time evolution and the error of extrapolating the signal, we choose to suppress parts of our original signal and the complete extrapolated signal. We exponentially suppress times after $t=40/3J$ with a Gaussian $w(t) = \exp(-2(t/t_0)^2)$ where $t_0 = 20/3J$. 

Finally, we perform a time Fourier transform into energy space which grants us access to the full ARPES spectrum $A_{\sigma}(\k, \omega)$.

\section{ARPES spectra}
\label{app:arpes_spectra}

\begin{figure*}[t]
    \centering
    \includegraphics[width = 0.95 \textwidth]{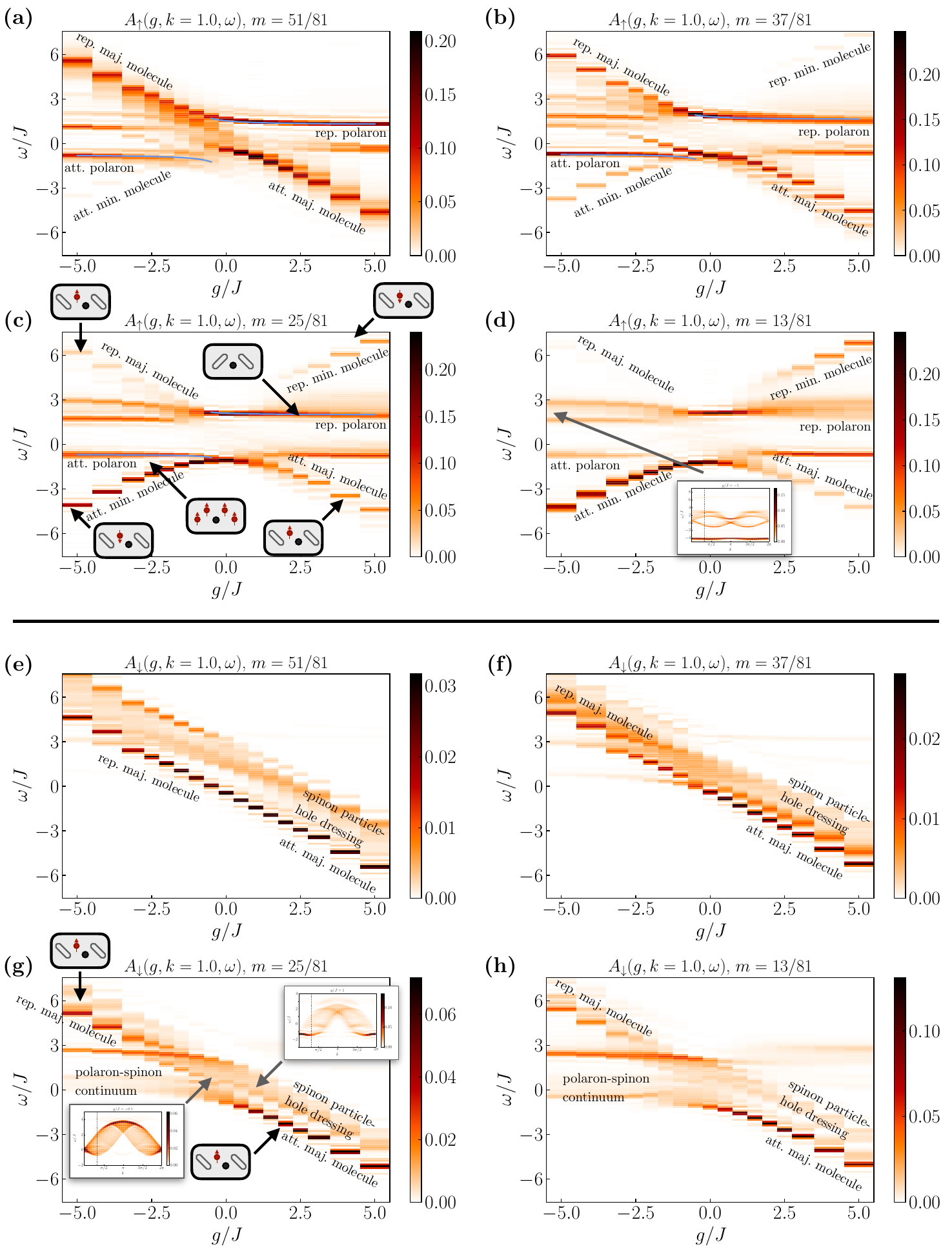}
    \caption{\textbf{ARPES spectra.} We plot the majority ($\sigma = \; \uparrow$, a-d) and minority ($\sigma = \; \downarrow$, e-h) ARPES spectrum Eq.~(\ref{eq:arpes}) of Hamiltonian~(\ref{eq:eMG}) at fixed GS momentum $k=1.0$ for $L=81$ and at different magnetizations $m$. We label the branches according to their qualitative behavior. We fit the polaron branches using the ansatz Eq.~(\ref{eq:fit_ansatz}) and plot the fits in blue. We find an almost flat band for the attractive minority molecule in the inset of (d).}
    \label{fig:overview}
\end{figure*}

In Fig.~\ref{fig:overview}, we show the interaction-dependent majority ($\sigma = \; \uparrow$, a-d) and minority ($\sigma = \; \downarrow$, e-h) ARPES spectra at fixed momentum $k=1.0$ at different magnetizations. We label branches according to their qualitative behavior and illustrate the fitting of the polaron branches (see below). Here we also briefly discuss the spectral weights of the polaron and the molecular branches.

The minority molecule is not visible in the minority spectrum because of its vanishing spectral weight: We remove a $\downarrow$-spinon from a GS where practically all $\downarrow$-spinons are bound in singlets, see Fig.~\ref{fig:overview}e-h. Conversely, the majority molecule ARPES weight increases for increasing $m$ since we have more free $\uparrow$-spinons in the GS. This is especially relevant for the majority spectrum where the entire majority molecule spectral weight originates from free $\uparrow$-spinons in the GS, see Fig.~\ref{fig:overview}a-d

The polaron branches are very dominant in the highly magnetized majority spectrum because of a high chance to remove a free $\uparrow$-spinon. Vice versa, in a highly magnetized minority spectrum, the removal of a down spin will break a singlet; the holon and the adjacent $\uparrow$-spinon form a molecule, resulting in a vanishing polaron spectral weight. 

We set $J=1$ for convenience and fit the attractive/repulsive polaron branches using the ansatz 
\begin{align} \label{eq:fit_ansatz}
    E_{\mathrm{a/r}}(g, m) = \frac{A_{\mathrm{a/r}}}{g -g_{\mathrm{c,a/r}}} \times m + B_{\mathrm{a/r}}(m),
\end{align}
where crucially $A_{\mathrm{a/r}}$ and $g_{\mathrm{c,a/r}}$ are independent of the magnetization. We find good agreement with $g_{\mathrm{c,a}} = 0.2$, $g_{\mathrm{c,r}} = -1.3$, and $A_{\mathrm{a}}=A_{\mathrm{r}} = 0.56$, the fits are the blue curves in Fig.~\ref{fig:overview}a-c. 

We can estimate the location of the Feshbach resonance as 
\begin{align}
    g_\mathrm{c} = \frac{g_{\mathrm{c,a}} + g_{\mathrm{c,r}}}{2} = -0.55,
\end{align}
which is in good agreement with Fig.~\ref{fig2}b. $g_\mathrm{eff}(g_c)$ can be estimated from Eq.~(\ref{eq:mean_field}) as
\begin{align}
    g_\mathrm{eff}(g_c) &= \frac{\langle \hat{\mathcal{H}}_{\mathrm{int}} \rangle}{m \times 
    \langle \hat{n}^{\mathrm{h}} \rangle} \nonumber \\
    &= \frac{ E_{\mathrm{r}}(g_\mathrm{c}, m) - E_{\mathrm{r}}(g \to \infty, m)}{m \times \langle \hat{n}^{\mathrm{h}} \rangle} \nonumber \\
    &\quad  \; \, - \frac{ E_{\mathrm{a}}(g_\mathrm{c}, m) - E_{\mathrm{a}}(g \to -\infty, m)}{m \times \langle \hat{n}^{\mathrm{h}} \rangle} \nonumber \\
    &= \frac{4 A_{\mathrm{a}}}{\bigl[ g_{ \mathrm{c,a}} - g_{\mathrm{c,r}} \bigr] \times \langle \hat{n}^{\mathrm{h}} \rangle} \nonumber \\
    &\approx 1.5 \approx 2.7 \times |g_{\mathrm{c}}|,
\end{align}
thus confirming that the effective interaction we find is Feshbach-like, i.e. resonantly enhanced! Intuitively, the interaction energy corresponds to the bending of the polaron branches which is a direct consequence of the resonantly enhanced interaction.

%

\end{document}